\documentclass{article}

\usepackage{PRIMEarxiv}
\usepackage{algorithm}
\usepackage{amsmath}
\usepackage{color,array,amsthm}
\usepackage{graphicx}
\usepackage{nicematrix}
\usepackage{amssymb}
\usepackage{algorithmic}
\usepackage{xcolor}
\usepackage{makecell}
\usepackage{colortbl}
\usepackage[final]{changes}
\usepackage{soul}
\usepackage{booktabs, multirow}
\usepackage[utf8]{inputenc} 
\usepackage[T1]{fontenc}    
\usepackage{hyperref}       
\usepackage{url}            
\usepackage{booktabs}       
\usepackage{amsfonts}       
\usepackage{nicefrac}       
\usepackage{microtype}      
\usepackage{lipsum}
\usepackage{fancyhdr}       
\usepackage{caption}
\title{A Drone-mounted magnetometer system for automatic interference removal and landmine detection}

\author{
  Alex Paul Hoffmann \\
  NASA Goddard Space Flight Center \\
  Greenbelt, Maryland\\
  alex.p.hoffmann@nasa.gov
  \And
  Matthew G. Finley \\
  University of Iowa \\
  Iowa City, Iowa \\
  matthew-g-finley@uiowa.edu
  \And
  Eftyhia Zesta \\
  NASA Goddard Space Flight Center \\
  Greenbelt, Maryland\\
  eftyhia.zesta@nasa.gov
  \And
  Mark B. Moldwin \\
  University of Michigan \\
  Ann Arbor, Michigan \\
  mmoldwin@umich.edu
  \And
  Lauro V. Ojeda \\
  University of Michigan \\
  Ann Arbor, Michigan \\
  lojeda@umich.edu
  }

\begin{document}
\maketitle

\begin{abstract}
Landmines have been extensively used in conflict zones as an indiscriminate weapon to control military movements, often remaining active long after hostilities have ended. Their presence poses a persistent danger to civilians, hindering post-war recovery efforts, causing injuries or death, and restricting access to essential land for agriculture and infrastructure. Unmanned aerial vehicles (UAV) equipped with magnetometers are commonly used to detect remnant hidden landmines but come with significant technical challenges due to magnetic field interference from UAV electronics such as motors. We propose the use of a frame-mounted UAV-borne two-magnetometer payload to perform a two-step automated interference removal and landmine detection analysis. The first step removes interference via the Wavelet-Adaptive Interference Cancellation for Underdetermined Platform (WAIC-UP) method designed for spaceflight magnetometers. The second method uses the Rapid Unsupervised Detection of Events (RUDE) algorithm to detect landmine signatures. This two-step WAIC-UP/RUDE approach with multiple magnetometers achieves high-fidelity ordinance detection at a low computational cost and simplifies the design of magnetic survey payloads. We validate the method through a Monte Carlo simulation of randomized landmine placements in a 10 x 10 m square grid and drone motor interference. Additionally, we assess the efficacy of the algorithm by varying the drone's altitude, examining its performance at different heights above the ground.
\end{abstract}


\section{Introduction}
Landmines continue to pose a severe threat in post-conflict regions, causing long-term humanitarian crises and impeding economic recovery. The escalation of the Russo-Ukrainian War in 2022 has led to the deployment of over two million landmines across Ukraine, a nation whose fertile soils are vital for Europe's agricultural output \cite{schindler}. These hidden hazards not only claim thousands of civilian lives each year but also render extensive areas of farmland unusable, significantly impacting food security and hindering economic development. The demining process is both time-consuming and prohibitively expensive; the United Nations estimates that clearing landmines in Ukraine could cost over \$37 billion \cite{worldbank}. This immense challenge underscores the urgent need for cost-effective and efficient landmine detection and clearance methods to restore safety and revitalize agricultural productivity in affected regions.

Detecting landmines is inherently challenging due to the \added{potential} use of non-magnetic materials in mine construction, the varying depths at which mines are buried, \added{and challenging terrain}. Unmanned aerial vehicles (UAVs) have emerged as valuable tools for surveying large areas for unexploded ordnance (UXO), equipped with an array of sensors such as ground-penetrating radar (GPR), cameras, and magnetometers. GPR operates by emitting electromagnetic waves into the ground and analyzing the reflected signals to identify anomalies indicative of landmines \cite{grathwohl, pisciotta}. While GPR is effective in detecting both metallic and non-metallic mines, its performance can be compromised by factors like soil moisture content and the depth of the buried mines \cite{robledo}. Visual analysis using machine learning can identify surface-laid or shallow-buried mines by detecting terrain changes \cite{buar}, but mines overgrown with vegetation and deeper mines pose challenges, requiring complex image processing techniques. Thermal imaging and acoustic sensing offer additional detection capabilities but are heavily dependent on environmental conditions, which can limit their reliability \cite{alqudsi}.

Magnetometers detect disturbances in the Earth's magnetic field caused by ferromagnetic materials, making them particularly useful for locating mines with \added{ferrous} content \cite{easymag, barnawi}. However, a significant challenge in using magnetometers mounted on UAVs is filtering out magnetic interference generated by the drone itself, which can severely affect detection accuracy. Traditional approaches to mitigate this interference involve mounting the magnetometers on towed platforms to increase their distance from the drone\added{; which limits the maneuverability of the UAV}, and applying low-pass filters to the magnetometer signals to remove high-frequency noise \cite{kolster, yoo}. Despite these efforts, such methods have limited success due to the presence of lower-frequency magnetic fields induced by the drone's motors \cite{walter}. Moreover, applying a low-pass filter reduces the spatial resolution of the magnetometer measurements, diminishing the accuracy of landmine position estimates. Recent work has attempted to apply the Sheinker and Moldwin Magnetic Gradiometry algorithm to UAV magnetometers for landmine detection \cite{sheinker, mu}; however, this algorithm assumes a single interference source and cannot adequately model the complex, near-field stray magnetic fields produced by UAVs. 

In contrast, significant advances have been made in spacecraft missions that study space plasmas, where removing spacecraft-generated magnetic fields \added{from subsystems such as solar panels and reaction wheels} is critical for accurate in situ measurements \cite{sheinker, mathen, magprime}. Building on these developments, we propose the use of Wavelet-Adaptive Interference Cancellation for Underdetermined Platforms (WAIC-UP) algorithm to remove drone interference from a pair of magnetometers mounted directly on the UAV \cite{waicup}. Originally designed for CubeSat magnetometers, WAIC-UP offers a cost-effective and computationally efficient solution that could significantly enhance magnetometer-based landmine detection from UAVs. WAIC-UP is an efficient algorithm for stray magnetic field removal from magnetometers in an unorthodox bus-mounted configuration \cite{magprime}, whereas the Sheinker and Moldwin algorithm \cite{sheinker, mu} requires a gradiometry configuration. Additionally, we use the Rapid Unsupervised Detection of Events (RUDE) algorithm to automatically identify landmines \cite{finley}. RUDE combines principal component analysis (PCA)  with unsupervised clustering through a \added{One-Class Support Vector Machine (OC-SVM)} to detect anomalies in the magnetometer time series data. The use of RUDE enables landmine detection without assumptions on the specific type of UXO present, and therefore decreases the burden of costly and time-intensive expert analysis in minefield clearing operations. \added{This approach is part of a broader magnetic anomaly detection (MAD) field of research for detecting ferromagnetic objects \cite{mad}. }

We validate the combined WAIC-UP and RUDE methodology through Monte Carlo simulations featuring varying drone interference signatures. In the first simulation, several landmines are randomly placed within a 100 m\textsuperscript{2} grid, and a simulated drone traverses the area in a sinusoidal pattern to search for mines. The second simulation places four landmines within the same grid and varies the drone's altitude from 50 cm to 3 m to assess the efficacy of the proposed method at different heights. These simulations allow us to statistically evaluate the proposed method against traditional low-pass filtering for interference reduction and hard-thresholding techniques for landmine detection.

The remainder of this paper is structured as follows. Section II provides an overview of the WAIC-UP and RUDE algorithms and details the formulation of the Monte Carlo simulations. In Section III, we present and analyze the simulation results. Section IV discusses the outcomes, underlying assumptions, and potential directions for future research. Finally, Section V summarizes our conclusions and highlights the impact of the proposed landmine detection algorithm on improving demining efforts.

\section{Methodology}
Stray magnetic field interference generated by UAV electrical currents inhibit \added{magnetic measurements} by UAV-mounted magnetometers. We consider a simulated UAV with two frame-mounted magnetometers. The WAIC-UP algorithm is applied to remove UAV stray magnetic field interference, and improve the signal-to-noise ratio of the magnetic footprint of the landmine. Next, the RUDE algorithm is applied to automatically detect the landmine signals in the cleaned time-series. This two-step method lowers the cost and technical complexity of using magnetometers for landmine detection and removal\added{, eliminating the need for unwieldy towed arrays or booms. Beyond landmine detection, this approach can also enhance the quality of magnetic data for other UAV-based magnetometer applications such as geophysical surveys.} 

\subsection{Interference removal through WAIC-UP}

To mitigate UAV-generated magnetic interference, we implement the WAIC-UP algorithm. Originally developed for spacecraft, WAIC-UP extends the Sheinker and Moldwin \cite{sheinker} gradiometry algorithm, which removes a single interference source using a pair of magnetometers. Our implementation adapts this approach for UAV applications, operating in the wavelet domain to account for multiple interference sources with varying spectral characteristics.

The key principle of WAIC-UP lies in leveraging the spectral difference between the landmine magnetic field and the UAV interference, making it well-suited for magnetometer payloads mounted directly on the UAV frame. The algorithm assumes two magnetometers, \( B_1(t) \) and \( B_2(t) \), record the mixed magnetic signals affected by both ambient magnetic fields and UAV interference. The system can be modeled as a linear mixing of two components: the landmine magnetic field signal, S$_{L}$(t), and the stray magnetic field signal from the UAV, S$_{UAV}$(t).

The time-domain relationship between the two magnetometer signals can be expressed as:

\begin{equation}\label{eq:1}
\begin{bmatrix}
B_{1}(t) \\ B_{2}(t) \\
\end{bmatrix}
=
\begin{bmatrix}
1 & 1  \\
1 & k  \\
\end{bmatrix}
\begin{bmatrix}
S_{L}(t)\\ S_{UAV}(t)\\ 
\end{bmatrix}
\end{equation}

where $k$ is a gain factor that accounts for the relative contribution of the interference to the second magnetometer. Here, S$_{L}$(t) represents the ambient magnetic field (containing the landmine magnetic fields) that we aim to recover, while S$_{UAV}$(t) denotes the UAV-generated stray magnetic field.

However, this time-domain system assumes that the UAV magnetic field is a single multipole field, and has limited ability to resolve multiple interference signals. To overcome this, we apply the continuous wavelet transform to both magnetometer signals, transforming them into the wavelet domain:

\begin{gather}\label{eq:2}
\begin{cases}
W_1(s) = X(s) + A(s) + \omega_1(s) \\
W_2(s) = X(s) + K(s)A(s) + \omega_2(s) 
\end{cases}
\end{gather}

where $W_1(s)$ and $W_2(s)$ are the wavelet transforms of $B_1(t)$ and $B_2(t)$, respectively, at scale $s$, with $X(s)$ representing the wavelet coefficients of the landmine magnetic field and $A(s)$ the wavelet coefficients of the UAV interference. The terms $\omega_1(s)$ and $\omega_2(s)$ account for noise in the magnetometer measurements, and K is a scale-dependent gain factor that defines the contribution of the interference signal at the second magnetometer.

To estimate the ambient field, we first calculate the difference between the two wavelet-transformed signals:

\begin{gather}\label{eq:3}
\begin{cases}
D(s) = W_2(s) - W_1(s) \\
D(s) = (K(s)-1)A(s) + \omega_2(s) - \omega_1(s) 
\end{cases}
\end{gather}

Given that the ambient/landmine magnetic field signal $X(s)$ is common to both magnetometers, the difference $D(s)$ isolates the contribution of the interference signal $A(s)$. Using a correlation-based estimator, we compute the gain factor $K$ for each wavelet scale s:

\begin{equation}\label{eq:4} \
\hat{K}(s) = \frac{\sum D(s) W_2(s)}{\sum D(s) W_1(s)} 
\end{equation}

With the estimated gain factor $\hat{K}(s)$, we reconstruct the landmine signal at each scale:

\begin{equation}\label{eq:5} 
\hat{X}(s) = \frac{\hat{K}(s) W_1(s) - W_2(s)}{\hat{K}(s) - 1} 
\end{equation}

By applying the inverse wavelet transform to the estimated $\hat{X}(s)$, we reconstruct the time-domain ambient magnetic field without UAV interference. This method enables the identification of multiple interference signals and allows for accurate and computationally efficient interference cancellation, enabling reliable detection of landmine magnetic signatures.

\subsection{Landmine detection through RUDE and HDBSCAN}
The Rapid Unsupervised Detection of Events (RUDE) algorithm~\cite{finley} is employed to automatically detect landmine signals via the identification of anomalies in the magnetic field measurements. RUDE leverages Dynamic Principal Component Analysis (D-PCA) to reduce the dimensionality of the measured time series data, and unsupervised clustering is then used to identify the most statistically significant features (i.e., anomalous data) while filtering out noise and redundant information. This approach allows RUDE to detect landmines without the need for prior information or expert-defined rules regarding the strength or spectral content of the unexploded ordnance's magnetic signature, making it versatile for real-world applications.

In RUDE, the original time series data is first restructured into a reduced trajectory matrix. This matrix, denoted as $X$, is constructed by concatenating consecutive intervals of the input time series into matrix form as:

\begin{equation}
X = 
\begin{bmatrix}
x(1) & x(L+1) & \cdots & x(RL - L + 1) \\
x(2) & x(L+2) & \cdots & x(RL - L + 2) \\
\vdots & \vdots & \ddots & \vdots \\
x(L) & x(2L) & \cdots & x(RL)
\end{bmatrix}
\end{equation}

where $R$ is the number of intervals and $L$ is the length of each interval. $X$ is then passed through a traditional PCA decomposition, and a low number of principal components are retained (in this manuscript's examples only two components are considered in subsequent steps). Using the reduced trajectory matrix as input to PCA provides greater information about the temporal structure of the time series, on the scale of the window length ($L$) during this decomposition step.

Next, the reduced-dimension (in this case, two-dimension, corresponding to the two components retained after PCA) representation of the original time-series data is passed through the unsupervised clustering provided by a One Class Support Vector Machine (OC-SVM). This clustering technique has seen application in a wide range of fields and shows a high degree of computational efficiency. The output of the OC-SVM is a binary label associated with each of the points produced by the decomposition step, identifying the point as anomalous or nominal. These labels are then inverted onto the original time series, and each sample in the input data is designated as anomalous or nominal. 

For this application, the RUDE algorithm is applied iteratively. For each sample in the time-series input data a confidence score is computed by normalizing the summation of the binary detections across a range of window lengths. This confidence score reflects anomaly likelihood instead of a binary anomaly detection. Samples where the confidence score exceeds a user-defined threshold are then selected for further analysis. To enhance the accuracy of landmine localization, the Hierarchical Density-Based Spatial Clustering of Applications with Noise (HDBSCAN) algorithm is applied to cluster the UAV positions at the selected points, grouping them based on their spatial proximity \cite{campello}. HDBSCAN efficiently clusters the UAV positions, filtering out noise and allowing the detection of the center of each landmine. Together, RUDE and HDBSCAN form a low-cost, computationally efficient framework for landmine detection and localization using UAV-mounted magnetometers.

\subsection{Monte Carlo benchmark design}
To evaluate the performance of the combined WAIC-UP and RUDE algorithms in mitigating UAV interference and detecting landmines, we designed two Monte Carlo benchmark simulations using Magpylib \cite{ornter} to model the magnetic fields generated by both landmines and UAVs. In addition to testing our algorithms, we \added{perform a baseline comparison} against low-pass filtering for interference removal and simple thresholding for landmine detection. Low-pass filtering attenuates high-frequency components of the signal to reduce noise and interference, while thresholding identifies anomalies by flagging signal values that exceed a predefined magnitude.

\begin{figure}[!ht]
    \centering
    \includegraphics[width=\linewidth]{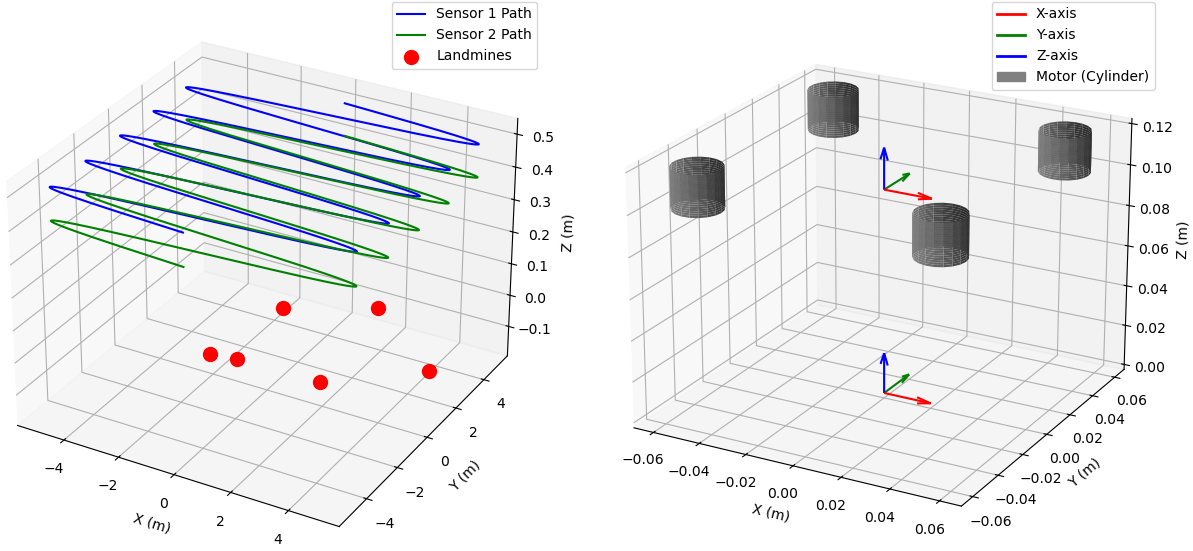}
    \caption{The left panel shows the UAV serpentine flight path over a 10 m x 10 m grid for landmine detection. The paths of the two magnetometers are shown in blue (Sensor 1) and green (Sensor 2). Red markers indicate landmine positions, which are randomly distributed and buried at depths between 0 to 15 cm. The right panel shows the layout of the UAV’s quadcopter motor setup. Each motor is displayed as a cylinder with a side length of 10 cm, with the magnetometers placed at the center of the UAV and 10 cm directly below the first magnetometer.}
    \label{fig:UAV}
\end{figure}

In the first benchmark, we simulated a 10 m $\times$ 10 m grid containing several M19 non-magnetic anti-tank mines. These mines have minimal metal content but still possess an appreciable magnetic footprint, characterized by a magnetic moment of \((-0.326,\,0.087,\,-0.338)\,\text{Am}^2\) \cite{lee}. We simulated these non-magnetic antitank mines to challenge the detection algorithms under difficult conditions. The landmines were randomly placed within the grid, maintaining a minimum separation of 2 meters between any two mines, and buried at depths ranging from 0 to 15 cm. A UAV equipped with two magnetometers traversed the grid following a serpentine flight path at a constant altitude of 50 cm and a speed of 1 m/s, resulting in a spatial resolution of approximately 1 cm per sample. \added{The entire flight path took 100 seconds to complete. Alternative speeds, sampling rates, and flight patterns can be chosen depending on the size of the area being surveyed. The serpentine flight pattern was chosen to yield a uniform coverage of measurements in the search grid. The choice of an optimal flight pattern in operations is subject to the constraints of the environment under survey.} Figure \ref{fig:UAV} illustrates the UAV design and an example landmine placement and UAV flight path.

The UAV was modeled as a quadcopter with four brushless permanent magnet synchronous (BPMS) motors arranged in a square configuration, with each side of the square measuring 10 cm. Two magnetometers were placed on the UAV and sampled at 100 Hz. The first magnetometer was positioned at the center of the square formed by the motors, in the same plane as the motors. The second magnetometer was mounted directly below the first, 10 cm beneath the plane of the motors. To simulate the inherent measurement uncertainty of the magnetometers, we added random normal noise with a 10 nT standard deviation to each measurement channel, reflecting the typical performance of low-cost magnetometers \cite{regoli}.

\begin{figure}[!htbp]
    \centering
    \includegraphics[width=0.5\linewidth]{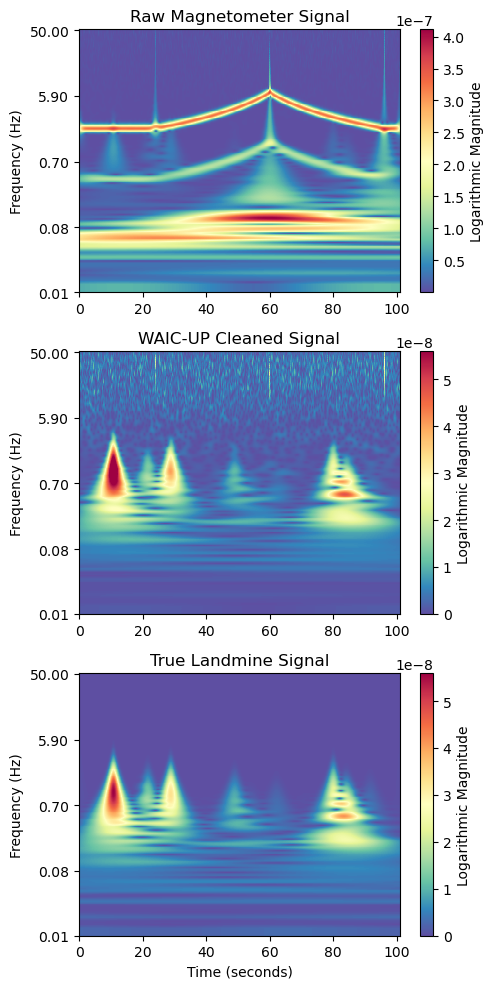}
    \caption{Example spectrogram of the simulated magnetic field signals. The top panel shows the raw magnetometer signal composed of the UAV interference and landmine signals. The middle panel shows the signal cleaned by WAIC-UP. The bottom panel shows the true landmine magnetic field signal with no UAV interference along the same path.}
    \label{fig:spec}
\end{figure}

To characterize the UAV's motor interference signals, we followed the model presented in \cite{walter}. Each motor's interference comprised three components: the magnetic field from the motor's mechanical rotation (F$_{MR}$), the rotating permanent magnet pairs (F$_{PM}$), and the induced magnetic field (F$_{IMF}$), with base frequencies set to 0.055 Hz, 0.39 Hz, and 2 Hz, respectively. To simulate variations in UAV motion such as acceleration and deceleration, the interference signals were randomly chirped by a factor between 1 and 5. Each motor component was simulated with dipole moments of 10 to 40 mA$\cdot$m$^2$ based on measurements by \cite{walter}. \added{At a distance of 10 cm from the motors, the magnetic field strength ranges from 2000 to 8000 nT. Additionally, these interference frequencies overlap with the spectral content of landmines, which increases the complexity of separating landmine signals from the interference.} Figure \ref{fig:spec} presents an example spectrograms of the raw magnetic field, the WAIC-UP cleaned signal, and the true ambient magnetic field for reference.

\newpage
In this benchmark, we conducted 250 randomized simulations, each containing between three and six landmines randomly positioned within the grid. The purpose was to create a set of diverse interference scenarios that enable the statistical evaluation of the WAIC-UP and RUDE algorithms compared to traditional methods. We evaluated four cases: (1) data cleaned using WAIC-UP with landmines detected by the RUDE algorithm; (2) data filtered with a standard 0.5 Hz low-pass filter and detection handled by RUDE; (3) data cleaned using WAIC-UP followed by simple thresholding for detection; and (4) data filtered with the low-pass filter followed by thresholding.

The second benchmark assessed the efficacy of the proposed algorithms with respect to UAV altitude. We conducted 35 simulations for each altitude, ranging from 50 cm to 3 m in 10 cm increments, while keeping the landmine positions fixed. This resulted in over a thousand simulations to evaluate the algorithms' performance at various heights. The limit for the thresholding detection method was set to 25 nT, which is 2.5 times the standard deviation of the measurement uncertainty.

For each simulation, we applied the HDBSCAN clustering algorithm \cite{campello} to the UAV's recorded positions where the RUDE algorithm or thresholding technique identified anomalies. Detections within 1 meter of a landmine were classified as true positives, while detections outside this range were considered false positives. Any landmines without an estimated detection within 1 meter were marked as false negatives. \added{For further clarification, a distribution of the distance between detections and the true landmine positions is plotted later in Fig. \ref{fig:dist}}

\section{Results}
\subsection{Benchmark 1: Random position simulations}
\begin{figure}
    \centering
    \includegraphics[width=0.5\linewidth]{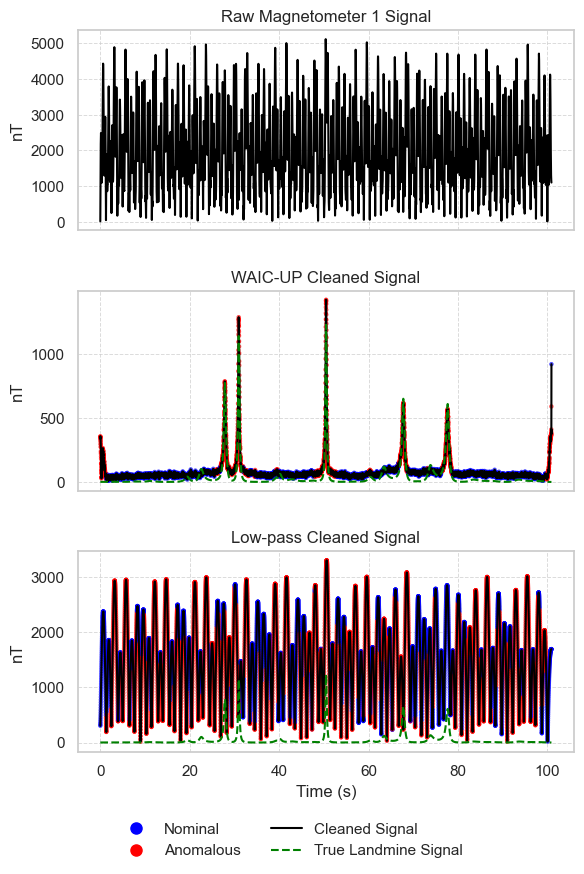}
    \caption{The top panel shows the raw absolute magnetic field signal from the bottom magnetometer in the UAV simulation. The middle panel shows the absolute magnetic field signal cleaned with the WAIC-UP algorithm. The bottom panel shows the absolute magnetic field signal cleaned with a simple low-pass filter. In the middle and bottom panel, the true landmine signal is plotted in the dashed green line. Anomalies identified by the RUDE algorithm are highlighted in red while nominal data is highlighted in blue.}
    \label{fig:time}
\end{figure}

Benchmark 1 evaluates the proposed interference removal and landmine detection algorithm through 250 randomized simulations. Each simulation has between 3 and 6 mines, simulated using the magnetic moment of the M19 minimum-metal anti-tank mine from \cite{lee}. \added{The M19 landmine has a large, square plastic body measuring 33 cm in diameter and 10 cm in height. The plastic casing houses only a few metal elements, including a thin stainless steel pressure plate, a copper detonator capsule, and a stainless steel firing pin.} In each simulation, either WAIC-UP or a low-pass filter was used to clean the raw magnetometer signals. After removing the noise, either the RUDE algorithm or thresholding was used to determine the presence of a landmine. The threshold for detection was set to 50 nT, and the confidence score threshold for detection with RUDE was set to 0.7. An example time-series that shows the WAIC-UP and low-pass filter results with the RUDE detection algorithm is show in Figure \ref{fig:time}. 

We mapped the landmine detections to the position of the UAV, and located them using the HDBSCAN clustering algorithm. From these located detections, we recorded the number of True Positives (TP), False Positives (FP), and False Negatives (FN) for each method of WAIC-UP / Low-pass and RUDE / Thresholding.

To evaluate the performance of the detection methods, we calculated the precision, recall, f1-score, and threat score \cite{ajithkumar}. Precision measures the proportion of detected landmines that are actually true landmines. It is defined as:

\begin{equation} \text{Precision} = \frac{\text{TP}}{\text{TP} + \text{FP}} \end{equation}

Recall (also known as Sensitivity) measures the proportion of actual landmines that were correctly detected. It is defined as:

\begin{equation} \text{Recall} = \frac{\text{TP}}{\text{TP} + \text{FN}} \end{equation}

F1-Score is the harmonic mean of Precision and Recall, providing a balance between the two. It is defined as:

\begin{equation} \text{F1-Score} = 2 \times \frac{\text{Precision} \times \text{Recall}}{\text{Precision} + \text{Recall}} \end{equation}

Threat Score (also known as Critical Success Index) measures the proportion of correct detections relative to all detections and missed detections, excluding true negatives. It is defined as:

\begin{equation} \text{Threat Score} = \frac{\text{TP}}{\text{TP} + \text{FN} + \text{FP}} \end{equation}

Additionally, we calculated the Pearson correlation coefficient $\rho$ between the cleaned magnetometer signals and true landmine magnetic field signal without UAV interference. The correlation coefficient is defined as:

\begin{equation} \rho = \frac{\sum (X_i - \overline{X})(Y_i - \overline{Y})}{\sqrt{\sum (X_i - \overline{X})^2 \sum (Y_i - \overline{Y})^2}} \end{equation}

where $X_i$ and $Y_i$ are the cleaned magnetometer signal and landmine magnetic field signals, respectively, and $\overline{X}$ and $\overline{Y}$ are their mean values.The \added{cumulative} results for each metric over the full 250 simulations are shown in Table \ref{table:landmine_detection_counts}.

\begin{table}[!h]
    \centering
    \begin{tabular}{|p{2cm}|p{2cm}|p{2.4cm}|p{2cm}|p{2.4cm}|}
    \hline
    \textbf{Metric} & \textbf{WAICUP/ RUDE} & \textbf{WAICUP/ Threshold} & \textbf{Lowpass/ RUDE} & \textbf{Lowpass/ Threshold} \\
    \hline
    \textbf{FP}  & \textbf{802} & 2446 & 2069 & 4293 \\
    \textbf{TP}   & 1028 & \textbf{1105} & 398 & 651 \\
    \textbf{FN}  & 87 & \textbf{10} & 717 & 464 \\
    \hline
    \textbf{Precision} & \textbf{0.562} & 0.311 & 0.161 & 0.132\\ 
    \textbf{Recall} & 0.922 & \textbf{0.991} & 0.357 & 0.584 \\ 
    \textbf{F1-Score} & \textbf{0.698} & 0.474 & 0.222 & 0.215\\ 
    \textbf{Threat Score} & \textbf{0.536} & 0.310 & 0.125 & 0.120\\
    \hline
    $\rho$-mean & \multicolumn{2}{|c|}{\textbf{0.9701}} & \multicolumn{2}{|c|}{0.0263} \\
    \hline
    \end{tabular}
    
    \caption{Detection metrics for each combination of interference removal and landmine detection methods. The bottom row contains the mean correlation for the low-pass and WAIC-UP techniques. The best algorithm for each metric is highlighted in bold-text.}
    \label{table:landmine_detection_counts}
\end{table}
\newpage

From the results in Table \ref{table:landmine_detection_counts} we observe that the combination of WAIC-UP interference removal and the RUDE detection algorithm achieved the highest overall performance. It attained a Precision of 0.562 and a Recall of 0.922, resulting in an F1-Score of 0.698 and a Threat Score of 0.536. This indicates that over 92\% of actual landmines were correctly detected, and more than half of the detections were true positives. WAIC-UP with thresholding has a higher Recall of 0.991, however, the false positive rate is three times larger. The high correlation coefficient between the WAIC-UP cleaned signal and the true landmine magnetic field signal of $\rho = 0.9701$ further demonstrates strong interference removal capabilities of the WAIC-UP algorithm over traditional low-pass filtering.

In contrast, using a low-pass filter for interference removal significantly reduced the performance of both detection methods. The Low-pass/RUDE combination yielded a Precision of 0.161 and a Recall of 0.357, while the Low-pass/Threshold method had the lowest Precision and F1-Score among all methods. These low performance metrics suggest that the low-pass filter was ineffective at removing UAV interference without distorting the landmine signals, leading to a high number of false positives and missed detections. \added{Furthermore, when we analyze the false negative results per simulation, we found that the combination of WAIC-UP with RUDE had no false negatives in 72\% of the simulations. The use of thresholding with WAIC-UP had a lower false negative rate with 96\% of the simulations having no false negative detections. The lowpass case for both thresholding and RUDE each had false negatives in over 90\% of the simulations.} Overall, the superior performance of the WAIC-UP methods, especially when combined with the RUDE algorithm, confirms the effectiveness of our proposed approach in preserving landmine signals while eliminating UAV-induced noise. This is critical in real-world demining operations, where both missed detections and false positives carry substantial risks.

\begin{figure}[ht!]
    \centering
    \includegraphics[width=0.7\linewidth]{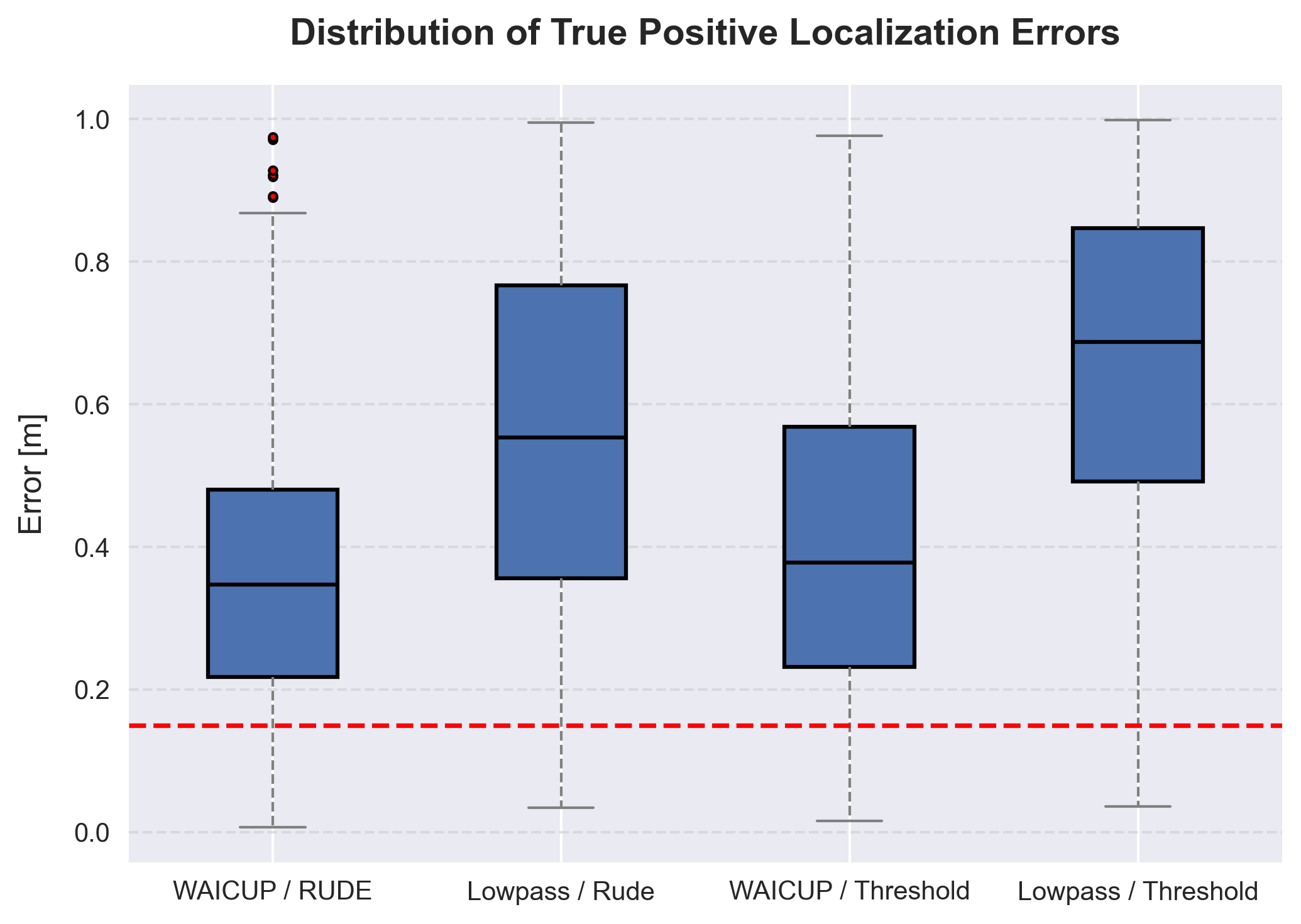}
    \caption{This box plot shows the error in meters between the estimated landmine positions and the actual center of the landmine for each interference removal and detection method. The red line shows the approximate radius of 15 cm of the M19 Landmine.}
    \label{fig:dist}
\end{figure}

For each true positive detection, we recorded the Euclidean distance between the estimated landmine position and the actual center of the landmine. Although the landmines were randomly buried at depths between 0 and 15 cm, we focused solely on the X-Y position and did not attempt to estimate the depth. The distribution of the distance from the estimated landmine position to the center of the landmine is shown in Figure \ref{fig:dist}.

The WAIC-UP combined with RUDE algorithm demonstrated the lowest median localization error of 36 cm, indicating superior accuracy. In contrast, the low-pass filter-based methods showed significantly higher median errors of 55 cm and 65 cm, reflecting less precise landmine localization. WAIC-UP's tighter distribution and fewer outliers further underscore its advantage in spatial resolution. In particular, these results suggest that the improved accuracy of WAIC-UP could facilitate safer demining operations, as the robotic mine excavator of \cite{Hemapala}, with a width of 87.6 cm, would benefit from more precise landmine positioning.

\subsection{Benchmark 2: Performance with respect to altitude.}
In the second benchmark, we evaluated the performance of the WAIC-UP and RUDE algorithms across different UAV altitudes, ranging from 50 cm to 3 m in 10 cm increments. Unlike the first benchmark, the landmines in these simulations were fixed at the corners of a 6 m x 6 m square, centered in the grid. For each altitude, we conducted 35 simulations using a seeded random number generator to ensure the same set of conditions at every height, resulting in over a thousand simulations in total. \added{This approach is useful for gauging performance with variable terrain or landmine depth, as the altitude changes simulate different real-world conditions.} The threshold for the hard thresholding detection method was set at 25 nT, while the confidence score threshold for RUDE was set to 0.70. The simulated UAV flew the same serpentine pattern as in Figure \ref{fig:UAV}.

\begin{figure}
    \centering
    \includegraphics[width=0.7\linewidth]{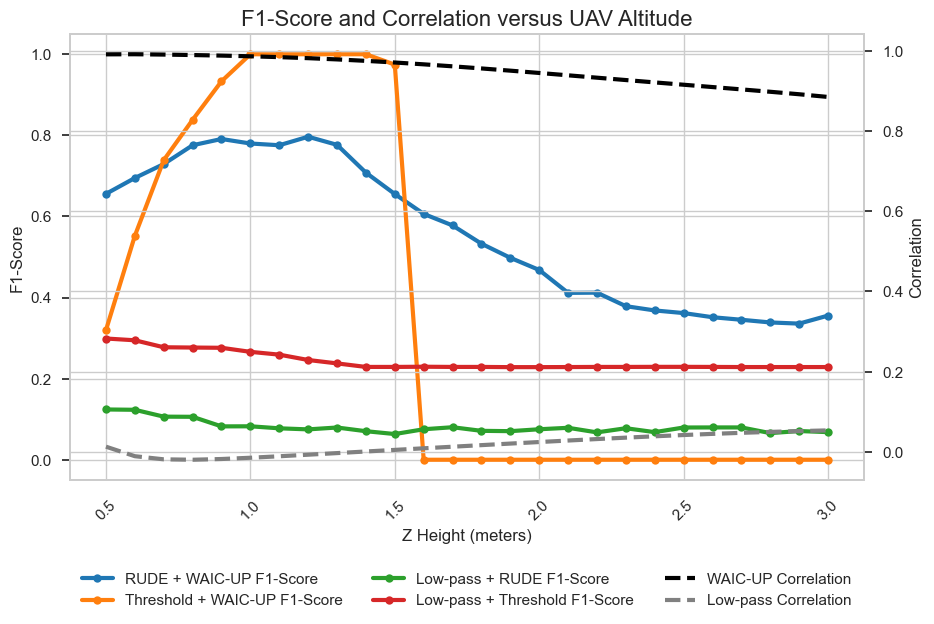}
    \caption{This plot illustrates the performance of different combinations of landmine detection algorithms in terms of F1-Score (left axis) and correlation (right axis) as a function of UAV altitude. Four detection methods are compared: WAIC-UP combined with RUDE, Low-pass filtering combined with RUDE, WAIC-UP with Thresholding, and Low-pass filtering with Thresholding.}
    \label{fig:altitude}
\end{figure}

We compared four combinations of interference removal and detection methods: WAIC-UP + RUDE, low-pass filtering + RUDE, WAIC-UP + thresholding, and low-pass filtering + thresholding. The results, shown in Figure \ref{fig:altitude}, indicate that both WAIC-UP-based methods outperformed the low-pass filter across all altitudes in terms of F1-Score and correlation to the true landmine signals.

WAIC-UP combined with RUDE achieved its highest F1-Score, peaking at approximately 0.80 between 0.7 and 1.3 meters altitude. However, performance steadily declined at higher altitudes, indicating reduced sensitivity at greater distances from the landmines. WAIC-UP combined with thresholding performed exceedingly well between 1 and 1.5 meters, where the F1-Score remained high, but exhibited a sharp drop-off after 1.4 meters. This drop occurred because the strength of the landmine signal at greater UAV altitudes fell below the 25 nT threshold, effectively zeroing out the detection performance beyond that range. This demonstrates that while WAIC-UP with thresholding is effective at moderate altitudes, it is particularly sensitive to signal strength, making it vulnerable to altitude increases or depth of burial.

In contrast, the low-pass filter-based methods performed consistently poorly, with both combinations maintaining an F1-Score below 0.4 across all altitudes. The Pearson correlation coefficient results further confirmed the superior performance of WAIC-UP, achieving near-perfect correlation at lower altitudes, while low-pass filtering struggled to remove interference adequately, resulting in poor correlation with the true landmine signal.

\section{Discussion}
The results of our simulations demonstrate that the combined use of the WAIC-UP interference removal algorithm and the RUDE detection method significantly outperforms traditional low-pass filtering and thresholding techniques in UAV-based landmine detection using magnetometers. The WAIC-UP algorithm effectively eliminates UAV-induced magnetic interference while preserving the subtle magnetic signatures of landmines, which is critical for accurate detection and localization. In Benchmark 1, our proposed method (WAIC-UP/RUDE) achieved a precision of 0.562 and a recall of 0.922, resulting in an F1-score of 0.698. This indicates that over 92\% of the actual landmines were correctly detected, and more than half of the total detections were true positives. In contrast, the low-pass filter combined with thresholding yielded a precision of 0.132 and a recall of 0.584, with an F1-score of only 0.215. The low-pass filtering approach struggled to differentiate between UAV interference and landmine signals due to the overlapping frequency components from the motors's mechanical rotation magnetic field \cite{walter}. This lead to a high number of false positives and missed detections.

The superior performance of the WAIC-UP algorithm can be attributed to its ability to operate in the time-frequency domain, effectively separating multiple interference sources with varying spectral characteristics. Unlike low-pass filtering, which uniformly attenuates high-frequency components (potentially removing essential landmine signals), WAIC-UP identifies the UAV magnetic interference signal and removes it, preserving the magnetic field signals of the landmines. Similarly, the RUDE detection algorithm outperformed simple thresholding by utilizing principal component analysis (PCA) and support vector machines (SVM) to identify anomalies in the magnetometer data. RUDE's machine learning approach allows it to adapt to the underlying statistical properties of the data, improving detection accuracy, especially when combined with effective interference removal.

Benchmark 2 further assessed the efficacy of our method at varying UAV altitudes. The WAIC-UP/RUDE combination maintained high detection performance up to an altitude of approximately 1.3 meters, with an F1-score peaking at 0.80. This demonstrates the method's robustness in real-world scenarios where UAV altitude may vary due to terrain or operational constraints. The decline in performance at higher altitudes is expected, as the magnetic signatures of landmines weaken with increased distance. However, even at reduced signal strengths, our method outperformed traditional techniques, which exhibited consistently low F1-scores across all altitudes.\added{ It is worth noting that older landmines, which typically contain more metal, generate stronger magnetic moments. As a result, detection at higher UAV altitudes may be more feasible for such mines compared to more modern designs with lower metal content.} On the other hand, the thresholding method with WAIC-UP performed perfectly from 1 to 1.5 m. However, thresholding relies on prior knowledge of the magnetic signature of the landmine, and changes with altitude. The RUDE algorithm, on the other hand, does not require any prior knowledge about the mine's magnetic signatures, burial depth, or UAV altitude.

Several assumptions underlie our simulations. First, the separation between the two magnetometers on the UAV must be an order of magnitude smaller than the distance to the landmine. This ensures that the system in eq. \ref{eq:1} is accurate and both sensors capture similar landmine signals while experiencing different levels of UAV interference. Second, we assumed that UAV interference is significantly larger than the intrinsic noise of the magnetometers. In situations where the interference is the same magnitude as the standard deviation of the measurement error, the WAIC-UP algorithm may not converge on the correct interference gain \cite{sheinker}.

Due to the nature of this study, there are a few limitations that need to be addressed. The WAIC-UP and RUDE algorithms are time-series algorithms and do not take into account the UAV position. The use of the HDBSCAN clustering algorithm for landmine localization may introduce errors, particularly in scenarios with closely spaced detections or unevenly spaced UAV flight paths. Future work could explore more sophisticated localization methods, such as incorporating machine learning algorithms that can handle complex spatial patterns and interpolating the magnetic field data into a grid \cite{mu}. \added{Moreover, these simulations did not account for other ferromagnetic objects, such as pipes, or detritus of war that may be present in the search area. These objects will be recognized as anomalies without further analysis. Other works have implemented an Euler deconvolution to filter magnetic field signals by their structural index \cite{mu}.} Lastly, our simulations only considered interference from brushless permanent magnet synchronous (BPMS) motors. In practice, UAVs may have other sources of magnetic interference, such as power cables or electronic components \cite{walter}. Expanding the interference model to include these factors would enhance the robustness of the interference removal algorithm.

For future research, testing the proposed method with real UAVs and actual landmines (or suitable surrogates) would provide valuable insights into its practical applicability and operational constraints. Field experiments could help validate the simulation results and identify any unforeseen challenges. Furthermore, improving the localization algorithm by integrating convolutional neural networks could enhance both the accuracy of landmine positioning and the ability to classify different types of ordnance based on their unique magnetic signatures.

\section{Conclusion}
In this study, we introduced a novel method for UAV-based landmine detection that leverages a dual magnetometer system mounted directly on the UAV frame. The application of the WAIC-UP interference removal algorithm effectively eliminates UAV-induced magnetic interference without the need for towed magnetometers, gradiometry configurations, or extended booms. By operating in the wavelet domain and accommodating multiple interference sources, WAIC-UP simplifies the magnetometer payload, making it more compact and easier to deploy. This advancement reduces the technical complexities associated with traditional magnetometer setups that use booms or towed arrays, and enables more efficient and flexible UAV operations in minefields. 

Furthermore, the integration of the RUDE detection algorithm automates the identification of landmines without requiring expert analysis or the development of complex magnetic field models. RUDE utilizes unsupervised machine learning techniques to detect anomalies in the magnetometer data, effectively distinguishing landmine signatures from background noise and interference. This automation reduces the reliance on specialized personnel and accelerates the detection process, making it more accessible and scalable for large-scale demining efforts.

By combining WAIC-UP and RUDE, we offer a low-cost and computationally efficient solution that enhances the accuracy and reliability of landmine detection using UAV-mounted magnetometers. This approach eliminates the need for bulky equipment and expert intervention, thereby reducing operational costs and complexities. Implementing this technology has the potential to significantly improve demining operations, facilitating the safe and rapid clearance of hazardous areas and contributing to the restoration of safe and productive lands in post-conflict regions.

\bibliographystyle{unsrt}

\end{document}